\documentclass[twocolumn,aps,prb,showpacs]{revtex4}

\usepackage{graphicx}
\usepackage{amsmath}
\usepackage{amssymb}

\begin{document}


\title{Superconductivity and magnetic fluctuations in electron-doped cobaltate superconductors}
\author{Ying Liang$^{1}$ and Bin Liu$^{2*}$}

\affiliation{$^{1}$Department of Physics, Beijing Normal University, Beijing 100875, China\\$^{2}$Department of Physics, Beijing Jiaotong University, Beijing 100044, China}

\begin{abstract}

We study the interplay between superconductivity and antiferromagnetic spin fluctuations in electron-doped cobaltate Na$_{x}$CoO$_{2}\cdot y$H$_{2}$O based on the kinetic energy driven superconductivity mechanism. We show that the superconducting state is governed by both charge carrier pairing and quasiparticle coherent, and displays a common dome-shaped phase diagram in agreement with experimental results. By calculating the dynamical spin structure factor, we theoretically find that the magnetic excitation shows a commensurate resonance peak, which locates at antiferromagnetic wave vector $Q (\frac{2\pi}{3},\frac{2\pi}{\sqrt{3}})$ for a broad range of low energies, then evolves outward into six incommensurate magnetic scattering peaks with increasing energy. Such commensurate-incommensurate spin resonance excitation should be measured by the inelastic neutron scattering technique (INS). Our present results strongly suggest that magnetic resonance can indeed be one of the fundamental features in doped Mott insulators.

\end{abstract}

\maketitle


\section{ Introduction}
In the past decade, superconductivity in sodium cobaltate Na$_{x}$CoO$_{2}\cdot y$H$_{2}$O (NCO) has been intensively studied in that it shares many similarities with high-T$_{c}$ cuprates\cite{takada,schaak,kastner}. The doped cobaltate has a lamellar structure consisting of two-dimensional (2D) triangular CoO$_{2}$ layers separated by a thick insulating layer of Na$^{+}$ ions and H$_{2}$O molecules\cite{takada}. Although the T$_{c}$ is a little bit lower, it displays the same kind of doping-controlled phase diagram that is observed in the doped cuprates\cite{takada,schaak}. In particular, the electron-doped NCO may be another unconventional superconductor coming from the family of doped Mott insulators. Although the ferromagnetic correlations is present in Na$_{x}$CoO$_{2}$ for large electron doping regimes\cite{motohashi}, the antiferromagnetic (AF) short-range spin correlations in Na$_{x}$CoO$_{2}$ and NCO at low doping levels has been observed from quadrupolar resonance and thermopower as well as other experimental measurements\cite{takada,schaak,uemura2,chu,wang}. Therefore, superconductivity mediated by the AF spin fluctuations in the doped cobaltate suggests the importance of the strong electron correlation as in the doped cuprates\cite{takada,schaak,kastner}. Understanding its superconducting (SC) mechanism could provide important insight into the microscopic origin of the unconventional superconductivity.

Recently, the kinetic energy driven SC mechanism applied into the doped cuprates based on the charge-spin separation fermion-spin theory\cite{feng2,feng1} has been successfully studied, where the dressed charge carriers interact occurring directly through the kinetic energy by exchanging spin excitations, leading to a net attractive force between dressed charge carriers, then the electron Cooper pairs originating from the dressed charge carriers pairing state are due to the charge-spin recombination, their condensation reveals the SC ground state, and the SC transition temperature is controlled by both charge carrier gap function and quasiparticle coherent weight. Since there is a remarkable resemblance in the normal and SC-state properties between NCO and the doped cuprates\cite{takada,schaak,kastner,chu} as mentioned above, and since the strong electron correlation is common for both materials, it thus could be expected that two systems may have similar underlying SC mechanism, i.e., superconductivity in the electron-doped cobaltate is also driven by the kinetic energy. Taking it into consideration, we in this paper study the unconventional superconductivity of sodium cobaltate NCO, and find that the SC state is governed by both charge carrier pairing and quasiparticle coherent, and the resulting SC transition temperature is suppressed to a lower temperature due to the strong magnetic frustration, which is in agreement with experimental measurements\cite{schaak}. Then we theoretically analyze the feedback effect of superconductivity with $d_{1}+ id_{2}$-wave pairing symmetry on the magnetic excitations, and find a commensurate resonance peak locating at antiferromagnetic wave vector $Q (\frac{2\pi}{3},\frac{2\pi}{\sqrt{3}})$ for a broad range of low energies, then evolving outward into six incommensurate magnetic scattering peaks with increasing energy, which is very similar to the case of electron-doped cuprates\cite{wilson,feng3,ilya}, and suggests that magnetic resonance could be intimately related to superconductivity. We propose that the inelastic neutron scattering experiment can verify the magnetic resonance excitations in electron-doped cobaltates NCO after the overcome of the difficulty in growing its high-quality SC sample.

\section{the $t$-$J$ Model and Kinetic Energy Driven SC Theory}

In electron-doped cobaltate, the characteristic feature is the presence of the 2D CoO$_{2}$ plane \cite{takada}, of which the essential physics is contained in the $t$-$J$ model on a triangular lattice \cite{baskaran,liu1} as
\begin{eqnarray}
H=-t\sum_{i\hat{\eta}\sigma}C_{i\sigma}^{\dagger}
C_{i+\hat{\eta}\sigma}-\mu\sum_{i\sigma}C_{i\sigma}^{\dagger}C_{i\sigma }+J\sum_{i\hat{\eta}}{\bf S}_{i}\cdot{\bf S}_{i+\hat{\eta}},
\end{eqnarray}
where $C^{\dagger}_{i\sigma}$($C_{i\sigma}$) is the fermion creation (annihilation) operator subject to the no-double-occupancy constraint,
${\bf S}_{i}=C^{\dagger}_{i}{\bf \sigma}C_{i}/2$ is the spin operator with ${\bf \sigma}=(\sigma_{x},\sigma_{y},\sigma_{z})$ as
the Pauli matrices, $\mu$ is the chemical potential. Note here that we work in the hole representation, so that a hole in the above model represents a physical electron-double-occupancy in electron-doped cobaltate.

We apply the charge-spin separation fermion-spin theory \cite{feng1}, within which the hole operators can be decoupled as $C_{i\uparrow}=a^{\dagger}_{i\uparrow}S^{-}_{i}$ and
$C_{i\downarrow}=a^{\dagger}_{i\downarrow}S^{+}_{i}$, where the spinful fermion operator $a_{i\sigma}=e^{-i\Phi_{i\sigma}}a_{i}$
describes the charge degree of freedom together with some effects of the spin configuration rearrangements, while $S_{i}$ describes the spin degree of freedom, then the single occupancy local constraint $\sum_{\sigma}
C^{\dagger}_{i\sigma}C_{i\sigma} =S^{+}_{i}a_{i\uparrow}
a^{\dagger}_{i\uparrow} S^{-}_{i}+ S^{-}_{i}a_{i\downarrow}
a^{\dagger}_{i\downarrow} S^{+}_{i}= a_{i}a^{\dagger}_{i} (S^{+}_{i}
S^{-}_{i}+S^{-}_{i} S^{+}_{i})=1- a^{\dagger}_{i} a_{i}\leq 1$ is satisfied in analytical calculations. Then the low-energy behavior of the $t$-$J$ model can be expressed as\cite{liu1,liu},
\begin{eqnarray}
H&=&-t\sum_{i\hat{\eta}}(a_{i\uparrow}S^{+}_{i}
a^{\dagger}_{i+\hat{\eta}\uparrow}S^{-}_{i+\hat{\eta}}+
a_{i\downarrow}S^{-}_{i}a^{\dagger}_{i+\hat{\eta}\downarrow}
S^{+}_{i+\hat{\eta}})\nonumber\\&-&\mu\sum_{i\sigma}a^{\dagger}_{i\sigma}
a_{i\sigma}+J_{{\rm eff}}\sum_{i\hat{\eta}}{\bf S}_{i}\cdot {\bf
S}_{i+\hat{\eta}},
\end{eqnarray}
with $J_{{\rm eff}}=(1-x)^{2}J$, and the doping level $x=\langle
a^{\dagger}_{i\sigma}a_{i\sigma}\rangle=\langle a^{\dagger}_{i}a_{i}\rangle$. In this case, the magnetic energy ($J$) term in $t$-$J$ model is only to form an adequate dressed spin configuration, while the kinetic energy ($t$) term has been transferred as the dressed charge carrier-spin interaction, which dominates the essential physics \cite{anderson1}. This dressed charge carrier-spin interaction is quite strong, and induces the dressed charge carrier pairing state (then the electron pairing state and superconductivity) by exchanging spin excitations in a higher power of the electron doping level as in the doped cupares\cite{feng2}. Then the order parameter for the electron Cooper pair without AF long range order (AFLRO) can be expressed as,
\begin{eqnarray}
\Delta=\langle C^{\dagger}_{i\uparrow}C^{\dagger}_{j\downarrow}-
C^{\dagger}_{i\downarrow}C^{\dagger}_{j\uparrow}\rangle=-\langle
S^{+}_{i}S^{-}_{j}\rangle\Delta_{a}
\end{eqnarray}
with the dressed charge carrier pairing order parameter $\Delta_{a}=\langle a_{j\downarrow} a_{i\uparrow}-
a_{j\uparrow}a_{i\downarrow}\rangle$. This shows that the dressed charge carrier pairs move freely in the background of the disordered spin
liquid state, and then the physical properties of SC state are essentially determined by the dressed charge carrier pairing state. Based on the Eliashberg's strong coupling theory \cite{eliashberg}, we obtain the self-consistent equations in terms of the method of equation of motion that is satisfied by the full dressed charge carrier diagonal and off-diagonal Green's functions $g(i-j,t-t') =\langle\langle
a_{i\sigma}(t);a^{\dagger}_{j\sigma}(t')\rangle \rangle$ and $\Im^{\dagger} (i-j,t-t')=\langle\langle
a^{\dagger}_{i\uparrow}(t); a^{\dagger}_{j\downarrow}(t') \rangle\rangle$ as
\begin{eqnarray}
g(k)=g^{(0)}(k)+g^{(0)}(k)
[\Sigma^{(a)}_{1}(k)g(k)-\Sigma^{(a)}_{2}
(-k)\Im^{\dagger}(k)], \\
\Im^{\dagger}(k)=g^{(0)}(-k)
[\Sigma^{(a)}_{1}(-k)\Im^{\dagger}(-k)+\Sigma^{*(a)}_{2} (-k)g(k)],
\end{eqnarray}
respectively, where the four vector notation $k=(i\omega_{n},\bf k)$, the mean-field (MF) dressed fermion Green's function $g^{(0)-1}(k) =i\omega_{n}-\xi_{\bf k}$, with the MF dressed charge carrier excitation spectrum $\xi_{\bf k}=Zt\chi\gamma_{\bf k}
-\mu$, $Z$ is the number of the nearest neighbor sites, $\gamma_{\bf k}=(1/Z)\sum_{\hat{\eta}}e^{i{\bf k}\cdot\hat{\eta}}=
[{\rm cos}k_{x}+2{\rm cos}(k_{x}/2){\rm cos}(\sqrt{3}k_{y}/2)]/3$, while the self-energies have been derived as,
\begin{eqnarray}
\Sigma^{(a)}_{1}(k)&=&(Zt)^{2}{1\over N^{2}}\sum_{{\bf
p,p'}}\gamma^{2}_{{\bf p+p'+k}}{1\over \beta}\sum_{ip_{m}}
g(p+k)\nonumber \\
&\times& {1\over\beta}\sum_{ip'_{m}}D^{(0)}(p')
D^{(0)}(p'+p), \\
\Sigma^{(a)}_{2}(k)&=&(Zt)^{2}{1\over N^{2}}\sum_{{\bf
p,p'}}\gamma^{2}_{{\bf p+p'+k}}{1\over \beta} \sum_{ip_{m}}\Im
(-p-k)\nonumber\\
&\times& {1\over\beta}\sum_{ip'_{m}}D^{(0)}(p')
D^{(0)}(p'+p),
\end{eqnarray}
where the MF dressed spin Green's function $D^{(0)}(i-j,t-t')=\langle\langle S^{+}_{i}(t);S^{-}_{j}(t')\rangle\rangle_{0}$.
Since the pairing force and dressed fermion gap function have been incorporated into the self-energy $\Sigma^{(a)}_{2}(k)$, then it
is an effective dressed charge carrier gap function. While self-energy $\Sigma^{(a)}_{1}(k)$ renormalizes the MF dressed charge carrier spectrum, so that it describes quasiparticle coherence and dominates the charge transport of the materials \cite{liu}. Moreover, $\Sigma^{(a)}_{2}(k)$ is an
even function of $i\omega_{n}$, while the self-energy $\Sigma^{(a)}_{1}(k)$ is not, which can be split into its symmetric and antisymmetric parts as $\Sigma^{(a)}_{1}(k)=\Sigma^{(a)}_{1e}(k)+i\omega_{n}\Sigma^{(a)}_{1o}(k)$ with $\Sigma^{(a)}_{1e}(k)$ and $\Sigma^{(a)}_{1o}(k)$ being even functions of $i\omega_{n}$. Now we define the charge carrier quasiparticle coherent weight $Z_{F}^{-1}(k) =1-\Sigma^{(a)}_{1o}(k)$.

We only study the static limit of the effective dressed charge carrier gap function and quasiparticle coherent weight, i.e., $\Sigma^{(a)}_{2}(k)=\bar{\Delta}_{a}({\bf k})$, and $Z_{F}^{-1}({\bf k}) =1-\Sigma^{(a)}_{1o}(\bf k)$. Then the dressed charge carrier diagonal and off-diagonal Green's functions are obtained as,
\begin{eqnarray}
g(k)&=&{1\over 2}\sum_{\nu=1,2}(1+{\bar{\xi_{\bf k}}\over
E_{\nu}(\bf k)}){Z_{F}({\bf k})\over i\omega_{n}-E_{\nu}(\bf k)},\\
\Im^{\dagger}(k)&=&-{1\over 2}\sum_{\nu=1,2}
{\bar{\Delta}^{*}_{aZ}({\bf k})\over E_{\nu}(\bf k)}{Z_{F}({\bf k})\over i\omega_{n}
-E_{\nu}(\bf k)},
\end{eqnarray}
with $E_{1}({\bf k})=E_{\bf k}$, $E_{2}({\bf k})=-E_{\bf k}$, $\bar{\xi_{\bf k}}=Z_{F}(\bf k)\xi_{\bf k}$, $\bar{\Delta}^{*}_{aZ}({\bf k})=Z_{F}({\bf k})\bar{\Delta}^{*}_{a}({\bf k})$ and the dressed fermion
quasiparticle spectrum $E_{\bf k}= \sqrt{\bar{\xi_{\bf k}}^{2}+ \mid
\bar{\Delta}_{aZ}(\bf k)\mid^{2}}$. Although $Z_{F}(\bf k)$ is a function of $\bf k$, the wave vector dependence is unimportant in that everything happens at the fermi surface (FS). In this case we approximate $Z_{F}(\bf k)$ by a constant $Z_{F}$ near the FS\cite{feng2}. So far the pairing symmetry in NCO is far from reaching a consensus, many experimental data point towards to spin-singlet Cooper pairing and suggest non-s-wave superconductivity without a full gap \cite{takada,schaak,chu,wang}. In particular, it has been argued according to the irreducible representations of the triangular lattice that there are three possible basis functions of even parity \cite{feng5}, i.e., one s-like function $s_{\bf k}={\rm cos} k_{x}+{\rm cos}[(k_{x}- \sqrt{3} k_{y})/2]+{\rm
cos} [(k_{x}+ \sqrt{3}k_{y})/2]$, and two d-like functions,
$d_{1\bf k}= 2{\rm cos}k_{x}-{\rm cos}[(k_{x}- \sqrt{3} k_{y})/2]-{\rm
cos}[(k_{x}+ \sqrt{3}k_{y})/2]$ and $d_{2\bf k}=\sqrt{3}{\rm cos}
[(k_{x}+\sqrt{3}k_{y})/2] -\sqrt{3}{\rm cos} [(k_{x}-
\sqrt{3}k_{y})/2]$. However, with the different linear combinations of these basis functions, it has been found
\cite{feng5} in terms of the Gutzwiller approximation scheme and variational Monte Carlo simulation that the lowest energy state is
the d-wave $(d_{1}+ id_{2})$ state. Therefore we in the following calculations consider both extended s-wave $\bar{\Delta}^{(s)}_{a}({\bf k})= \bar{\Delta}^{(s)}_{a}s_{\bf k}$ and the d-wave case $\bar{\Delta}^{(d)}_{a}({\bf k})= \bar{\Delta}^{(d)}_{a} (d_{1\bf k}+id_{2\bf k})$, and then the effective dressed fermion gap parameter and quasiparticle coherent weight satisfy the following equations,
\begin{eqnarray}
1&=&(Zt)^{2}{1\over N^{3}}\sum_{{\bf k,q,p}}\gamma^{2}_{{\bf k+q}}
\gamma^{(\tau)}_{{\bf k-p+q}}\gamma^{(\tau)}_{{\bf k}}{Z^{2}_{F}\over
E_{{\bf k}}}{B_{{\bf q}}B_{{\bf p}}\over\omega_{{\bf q}}
\omega_{{\bf p}}} \nonumber \\
&\times& \left({F^{(1)}_{1}({\bf k,q,p})\over (\omega_{{\bf p}}-
\omega_{{\bf q}})^{2}-E^{2}_{{\bf k}}} +{F^{(2)}_{1}({\bf k,q,p})
\over (\omega_{{\bf p}}+\omega_{{\bf q}})^{2}-E^{2}_{{\bf k}}}
\right ) ,\\
Z^{-1}_{F}&=&1+(Zt)^{2}{1\over N^{2}}\sum_{{\bf q,p}}
\gamma^{2}_{{\bf p}}Z_{F}{B_{{\bf q}}B_{{\bf p}}\over
4\omega_{{\bf q}}\omega_{{\bf p}}} \nonumber \\
&\times& \left({F^{(1)}_{2}({\bf q,p}) \over (\omega_{{\bf p}}
-\omega_{{\bf q}}-E_{{\bf p-q}})^{2}}\right . +{F^{(2)}_{2}({\bf q,p})\over (\omega_{{\bf p}}- \omega_{{\bf
q}} +E_{{\bf p-q}})^{2}} \nonumber \\
&+&{F^{(3)}_{2}({\bf q,p}) \over (\omega_{{\bf p}}+ \omega_{{\bf
q}}-E_{{\bf p-q}} )^{2}}+\left . {F^{(4)}_{2}({\bf q,p})\over (\omega_{{\bf
p}}+\omega_{{\bf q}}+E_{{\bf p-q}})^{2}} \right ),
\end{eqnarray}
respectively, where $\tau={\rm s,d}$, $F^{(1)}_{1}({\bf k,q,p})=
(\omega_{{\bf p}}- \omega_{{\bf q}})[n_{B}(\omega_{{\bf q}})-
n_{B}(\omega_{{\bf p}})] [1-2 n_{F}(E_{{\bf k}})]+E_{{\bf k}}
[n_{B}(\omega_{{\bf p}}) n_{B}( -\omega_{{\bf q}})+
n_{B}(\omega_{{\bf q}}) n_{B}(-\omega_{{\bf p}}) ]$,
$F^{(2)}_{1}({\bf k,q,p}) =-(\omega_{{\bf p }}+\omega_{{\bf q}})
[n_{B}(\omega_{{\bf q}}) -n_{B}(-\omega_{{\bf p}})][1-2 n_{F}
(E_{{\bf k}})]+E_{{\bf k}} [n_{B}(\omega_{{\bf p}}) n_{B}
(\omega_{{\bf q}}) +n_{B}(-\omega_{{\bf p}})n_{B}(-\omega_{{\bf q}
})]$, $F^{(1)}_{2}({\bf q,p})=n_{F}(E_{{\bf p- q+k_{0}}})[n_{B}
(\omega_{{\bf q}})-n_{B}(\omega_{{\bf p}})]- n_{B}(\omega_{{\bf p}
})n_{B}(-\omega_{{\bf q}})$, $F^{(2)}_{2} ({\bf q,p})= n_{F}
(E_{{\bf p-q+k_{0}}}) [n_{B}(\omega_{{\bf p}})-n_{B} (\omega_{{\bf
q}})]-n_{B}(\omega_{{\bf q}})n_{B} (-\omega_{{\bf p}})$,
$F^{(3)}_{2}({\bf q,p})= n_{F}(E_{{\bf p-q+k_{0}}})
[n_{B}(\omega_{{\bf q}})-n_{B}(-\omega_{{\bf p}})]+n_{B}
(\omega_{{\bf p}})n_{B}(\omega_{{\bf q}})$, and $F^{(4)}_{2}({\bf
q,p})=n_{F}(E_{{\bf p-q+k_{0}}})[n_{B}(-\omega_{{\bf q}})-n_{B}
(\omega_{{\bf p}})]+n_{B}(-\omega_{{\bf p}})n_{B}(-\omega_{{\bf
q}})$. These two equations determine the SC order directly, and must be solved simultaneously with other self-consistent equations, then all order parameters are determined by the self-consistent calculation
\cite{feng1}. In terms of the off-diagonal Green's function (9), we obtain the dressed charge carrier pair order parameter as,
\begin{eqnarray}
\Delta^{(\tau)}_{a}={2\over N}\sum_{k}\mid\gamma^{(\tau)}_{\bf k}\mid^{2}
{Z_{F}\bar{\Delta}^{(\tau)}_{aZ}\over E_{\bf k}}{\rm tanh} [{1\over 2}\beta E_{\bf
k}].
\end{eqnarray}
The dressed charge carrier pairing state originating from the kinetic energy term by exchanging dressed spin excitations will also
lead to form the electron Cooper pairing state as mentioned in Eq. (3). For a discussion of the physical properties in SC state,
we now need to calculate the electron off-diagonal Green's function $\Gamma^{\dagger}(i-j,t-t')=\langle\langle
C^{\dagger}_{i\uparrow}(t);C^{\dagger}_{j\downarrow}(t')\rangle\rangle$. In the framework of the charge-spin separation
fermion-spin theory\cite{feng1}, it is a convolution of the dressed spin Green's function $D^{(0)}(p)$ and off-diagonal dressed charge carrier Green's function $\Im(k)$, and can be expressed as,
\begin{eqnarray}
\Gamma^{\dagger}(k)&=&{1\over N}\sum_{p}{1\over \beta}
\sum_{ip_{m}}D^{(0)}(p)\Im(p-k)\nonumber\\&=&{1\over N}\sum_{{\bf p}} {Z_{F}
\bar{\Delta}^{(\tau)}_{aZ}({\bf p-k})\over E_{{\bf p-k}}}{B_{{\bf p}}
\over 2\omega_{{\bf p}}}\nonumber\\
&\times&\left ({(\omega_{{\bf p}}+E_{{\bf p-k}})
[n_{B}(\omega_{{\bf p}})+n_{F}(-E_{{\bf p-k}})] \over
(i\omega_{n})^{2}-(\omega_{{\bf p}}+E_{{\bf p-k}})^{2}}
\right . \nonumber \\
&-& \left . {(\omega_{{\bf p}}-E_{{\bf p-k}})[n_{B}(\omega_{{\bf
p}})+n_{F}(E_{{\bf p-k}})]\over (i\omega_{n})^{2}-(\omega_{{\bf
p}}-E_{{\bf p-k}})^{2}} \right ),
\end{eqnarray}
then the SC gap function is obtained as,
\begin{eqnarray}
\Delta^{(\tau)}({\bf k})&=&-{1\over \beta}\sum_{i\omega_{n}}
\Gamma^{\dagger}({\bf k},i\omega_{n})\nonumber \\
&=&-{1\over N}\sum_{{\bf p}} {Z_{F}\bar{\Delta}^{(\tau)}_{aZ}({\bf
p-k})\over 2E_{{\bf p-k}}}{\rm tanh}[{1\over 2}\beta E_{{\bf
p-k}}]\nonumber \\
&\times&{B_{{\bf p}}\over 2 \omega_{{\bf p}}}{\rm coth}[{1\over
2}\beta\omega_{{\bf p}}],
\end{eqnarray}which can be further written as $\Delta^{(\tau)}({\bf k})=\Delta^{(\tau)}\gamma_{\bf k}^{\tau}$. With the help of Eqs. (14) and (12), the SC gap
\begin{figure}[ht]
\includegraphics[scale=0.45]{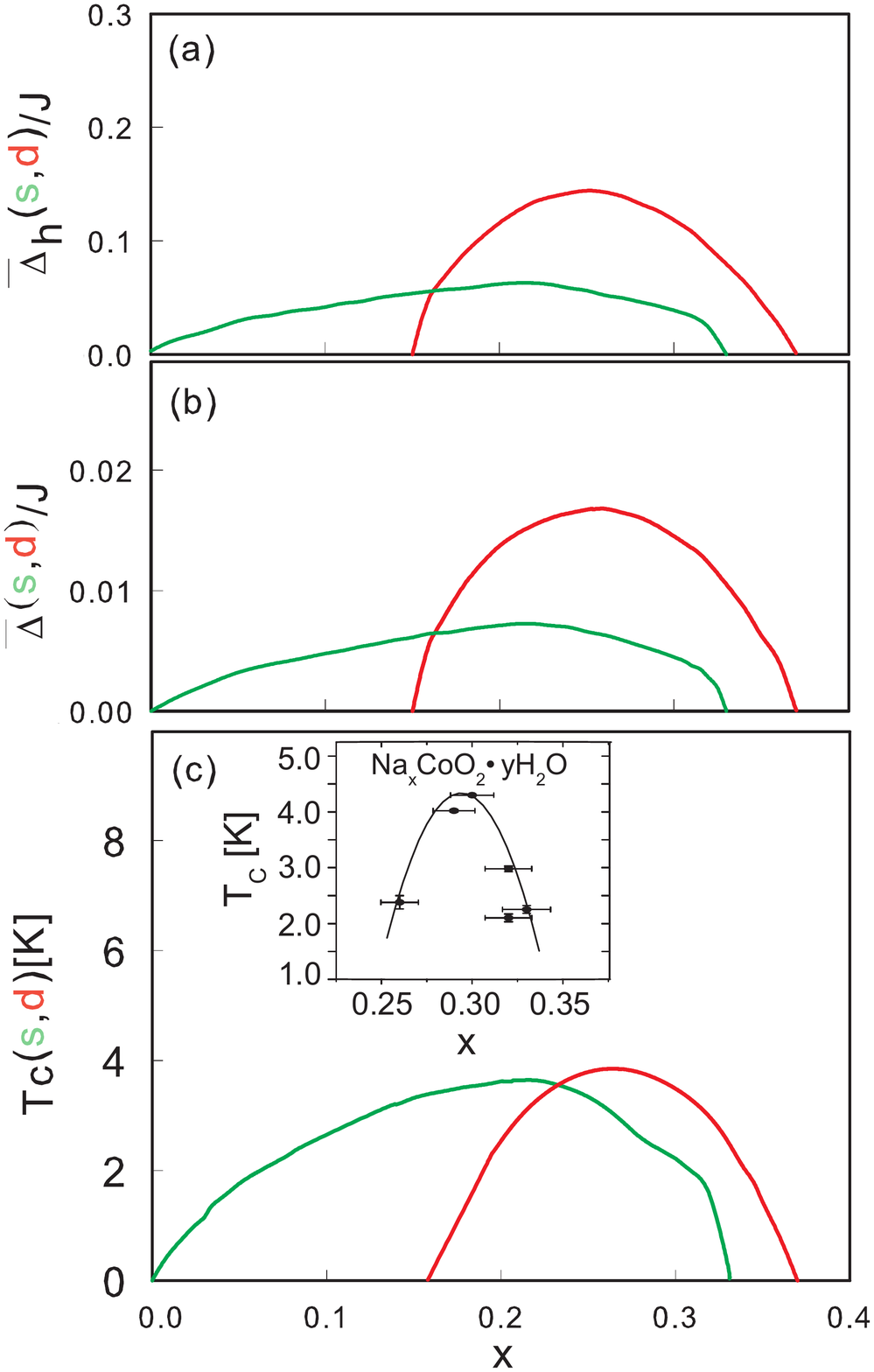}
\caption{(Color online) The
effective dressed charge carrier pairing (a) and effective superconducting
(b) gap parameters in the s-wave symmetry (green line) and d-wave
symmetry (red line) as a function of doping
concentration in $T=0.002J$ and $t/J=-3.0$. The
superconducting transition temperature (c) as a function of doping concentration in the s-wave symmetry (green line) and
d-wave symmetry (red line). Inset: the
experimental result taken from Ref. [2].}
\end{figure}
parameter can be obtained as $\Delta^{(\tau)}=-\chi\Delta^{(\tau)}_{a}$, then the effective SC gap parameter should be $\bar{\Delta}^{(\tau)}\sim-\chi\bar{\Delta}^{(\tau)}_{a}$. As in the doped cuprates \cite{feng2}, our present theory clearly indicates that there is a coexistence of the electron Cooper pair and short-range AF correlation in the doped cobaltates, and
therefore the short-range AF fluctuation can persist into superconductivity. The reason is that the AF fluctuation is
dominated by the scattering of dressed spins \cite{feng1}, which has been incorporated into the electron off-diagonal Green's
function (and hence the electron Cooper pair) in terms of the dressed spin Green's function. This result is consistent with some experimental results \cite{takada,schaak,uemura2,chu,wang}.

We perform a numerical calculation for the gap parameters, and the
results of the dressed charge carrier (a) and SC (b) gap parameters in the s-/d-wave symmetry as a function of the electron doping $x$ at $T=0.001J$ and $t/J=-3.0$ are shown in Fig. 1, where both values of the dressed charge carrier and SC gap parameters increase with increasing doping in the underdoped regime, and reach the maximal value for a particular doping concentration, then decrease in the overdoped regime. Although there is a coexistence of the electron Cooper pair and short-range AF correlation, the value of the SC gap parameter is still
suppressed by this AF fluctuation.

The SC transition temperature $T_{c}$ occurs when $\Delta^{(\tau)}=0$, which is identical to the dressed charge carrier pair
transition temperature occurring in the case of $\bar{\Delta}^{(\tau)}_{a}=0$. The SC transition temperature $T_{c}$
as a function of doping $x$ in the s-/d-wave symmetry for $t/J=-3.0$ is plotted in Fig. 1 (c) in comparison
with the experimental data \cite{schaak} taken from NCO (inset). For the s-wave symmetry, the maximal SC transition temperature T$^{(s)}_{c}$ occurs around a particular doping concentration $x\approx 0.21$, and then decreases in both lower doped and higher doped regimes.
However, for the d-wave symmetry, the maximal SC transition temperature T$^{(d)}_{c}$ occurs around the optimal doping
concentration $x_{{\rm opt}}\approx 0.29$, and then decreases in both underdoped and overdoped regimes. It is clearly shown that the SC state has the d-wave symmetry in a wide range of doping, in qualitative agreement with the experimental data \cite{schaak,uemura2}, supporting the domination of $d_{1}+ id_{2}$ pairing symmetry in NCO. Using an reasonable estimation value of $J\sim$ 15 mev to 20 mev in NCO, the SC transition temperature in the optimal doping is T$^{{\rm optimal}}_{c}\approx 0.02J\approx4{\rm K}$\cite{baskaran,liu1,liang}. Since NCO is the electron doped Mott insulator on a triangular lattice, the system has strong geometrical spin frustration. This magnetic frustration also induces the strong charged carrier's quantum fluctuation. In comparison with the case in the doped cuprates, we find that the
SC transition temperature in the doped cobaltates is suppressed heavily to a lower temperature due to both strong magnetic
frustration and dressed charge carrier's quantum fluctuation.

\section{Energy Dependent Commensurate and Incommensurate Magnetic Resonance}

We now turn to discuss the feedback effect of superconductivity with the $d_{1}+ id_{2}$-wave pairing symmetry on the magnetic excitations in the doped cobaltates. Among various phenomena of the spin excitations in SC state, the neutron spin resonance or the magnetic resonance observed in SC cuprates is undoubtly the most important feature, which has been regarded as an essential feature in doped Mott insulators. Unfortunately, due to the difficulty in growing the high-quality NCO SC sample there is no inelastic neutron scattering experiments on this system, so that it will be timely and necessary to study this important issue from the point of theoretical view.

Within the kinetic energy driven SC mechanism, the AF fluctuation is
dominated by the scattering of dressed spins with the full dressed spin Green's function
\begin{eqnarray}
D({\bf k},\omega)={1\over D^{(0)-1}({\bf k},\omega)-\Sigma^{(s)}
({\bf k},\omega)},
\end{eqnarray}where the second order spin self-energy $\Sigma^{(s)}({\bf k},\omega)$ is obtained from the dressed charge carrier bubble in terms of the collective mode in the dressed charge carrier particle-particle channel as,
\begin{eqnarray}
\Sigma^{(s)}({\bf k},\omega)&=&(Zt)^{2}{1\over N^{2}}\sum_{{\bf
p,q}}(\gamma^{2}_{{\bf q+p+k}}+\gamma^{2}_{{\bf p-k}})\nonumber\\
&\times&{B_{{\bf q+k}}\over\omega_{{\bf q+k}}}{Z^{2}_{F}\over
4}{\bar{\Delta}^{(d)}_{aZ}({\bf p}) \bar{\Delta}^{(d)}_{aZ}({\bf
p+q})\over E_{{\bf p}}E_{{\bf p+q}}} \nonumber \\
&\times& \left ( {F^{(1)}_{s}({\bf k,p,q})\over \omega^{2}-
(E_{{\bf p}} -E_{{\bf p+q}}+\omega_{{\bf q+k}})^{2}} \right .
\nonumber \\
&+&{F^{(2)}_{s}({\bf k,p,q})\over \omega^{2}-(E_{{\bf p+q}}
-E_{{\bf p}}+\omega_{{\bf q+k}})^{2}} \nonumber \\
&+& {F^{(3)}_{s}({\bf k,p,q})\over \omega^{2}-(E_{{\bf p}} +
E_{{\bf p+q}} +\omega_{{\bf q+k}})^{2}} \nonumber \\
&+& \left . {F^{(4)}_{s}({\bf k,p,q})\over \omega^{2}- (E_{{\bf
p+q}}+E_{{\bf p}}-\omega_{{\bf q+k}})^{2}} \right ),
\end{eqnarray}where $F^{(1)}_{s}({\bf k,p,q})=(E_{{\bf p}}-E_{{\bf p+q}}+
\omega_{{\bf q+k}})\{n_{B}(\omega_{{\bf q+k}})[n_{F}(E_{{\bf
p}})-n_{F}(E_{{\bf p+q}})]-n_{F}(E_{{\bf p+q}})n_{F}(-E_{{\bf
p}})\}$, $F^{(2)}_{s}({\bf k,p,q})=(E_{{\bf p+q}}-E_{{\bf p}}+
\omega_{{\bf q+k}})\{n_{B} (\omega_{{\bf q+k}})[n_{F}(E_{{\bf
p+q}})-n_{F} (E_{{\bf p}})]-n_{F}(E_{{\bf p}})n_{F}(-E_{{\bf
p+q}})\}$, $F^{(3)}_{s}({\bf k,p,q}) =(E_{{\bf p}}+E_{{\bf p+q}}
+\omega_{{\bf q+k}})\{n_{B} (\omega_{{\bf q+k}})[n_{F}(-E_{{\bf
p}})- n_{F}(E_{{\bf p+q}})] +n_{F}(-E_{{\bf p+q}})n_{F}(-E_{{\bf
p}})\}$, $F^{(4)}_{s}({\bf k,p,q})=(E_{{\bf p}}+E_{{\bf p+q}}-
\omega_{{\bf q+k}})\{n_{B} (\omega_{{\bf q+k}})[n_{F}(-E_{{\bf
p}})-n_{F}(E_{{\bf p+q}})] -n_{F}(E_{{\bf p+q}})n_{F}(E_{{\bf
p}})\}$. Then the dynamical spin structure factor which determines the spin excitations in the SC state can be written as,
\begin{eqnarray}
&S({\bf k},\omega)&=-2[1+n_{B}(\omega)]{\rm Im}D({\bf k},\omega)\nonumber \\
&=&{2[1+n_{B}(\omega)]B^{2}_{{\bf k}}{\rm Im} \Sigma^{(s)}({\bf k},\omega)
\over [\omega^{2}-\omega^{2}_{{\bf k}}-B_{{\bf k}}{\rm Re}
\Sigma^{(s)}({\bf k},\omega)]^{2}+[B_{{\bf k}}{\rm Im}
\Sigma^{(s)}({\bf k},\omega)]^{2}},
\end{eqnarray}where ${\rm Im}\Sigma^{(s)}({\bf k},\omega)$ and ${\rm Re}\Sigma^{(s)}({\bf k},\omega)$ are imaginary and real parts of
the second order self-energy respectively. It is shown that the above dynamical spin structure
factor has a well-defined resonance character, where $S({\bf k},\omega)$ exhibits resonance peaks when the incoming neutron
energy $\omega$ is equal to the renormalized spin excitation, i.e., $\omega^{2}-\omega_{{\bf k}_{c}}^{2}
-B_{{\bf k}_{c}}{\rm Re}\Sigma^{(s)}({\bf k}_{c}, \omega)=0$ for certain critical wave vectors ${\bf k}_{c}$, then the weight of these peaks is dominated by the inverse of the imaginary part of the dressed spin self-energy $1/{\rm Im}\Sigma^{(s)}({\bf k}_{c},\omega)$.
\begin{figure}[ht]
\includegraphics[scale=0.45]{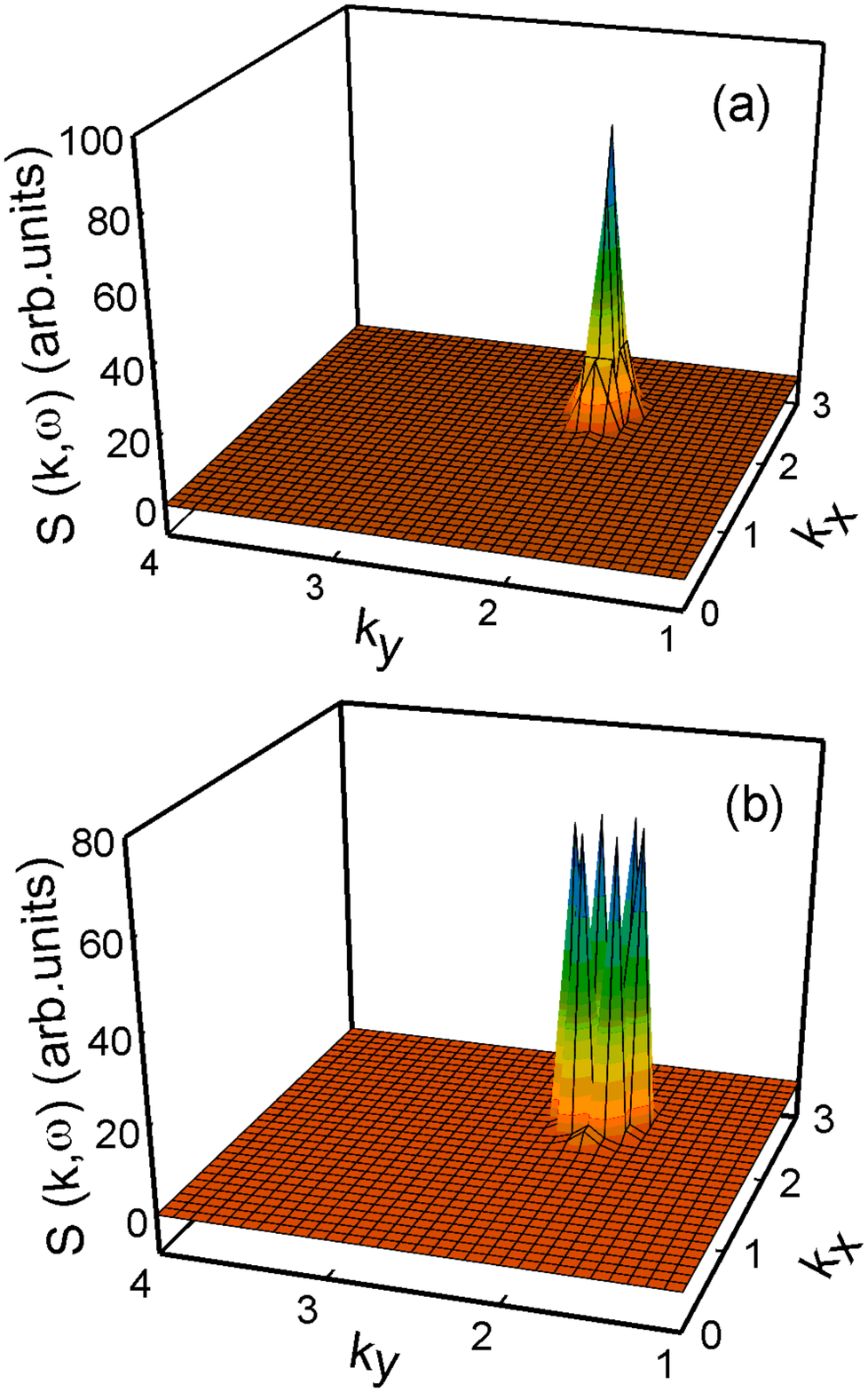}
\caption{(color online) The
dynamical spin structure factor $S({\bf k},\omega)$ in the
($k_{x},k_{y}$) plane as a unit of [$\pi/3,\pi/\sqrt{3}$] at $x_{{\rm opt}}=0.29$ with $T=0.002J$ for
$t/J=-3.0$ at (a) $\omega =0.08J$, and (b)
$\omega =0.20J$.}
\end{figure}
\begin{figure}[ht]
\includegraphics[scale=0.45]{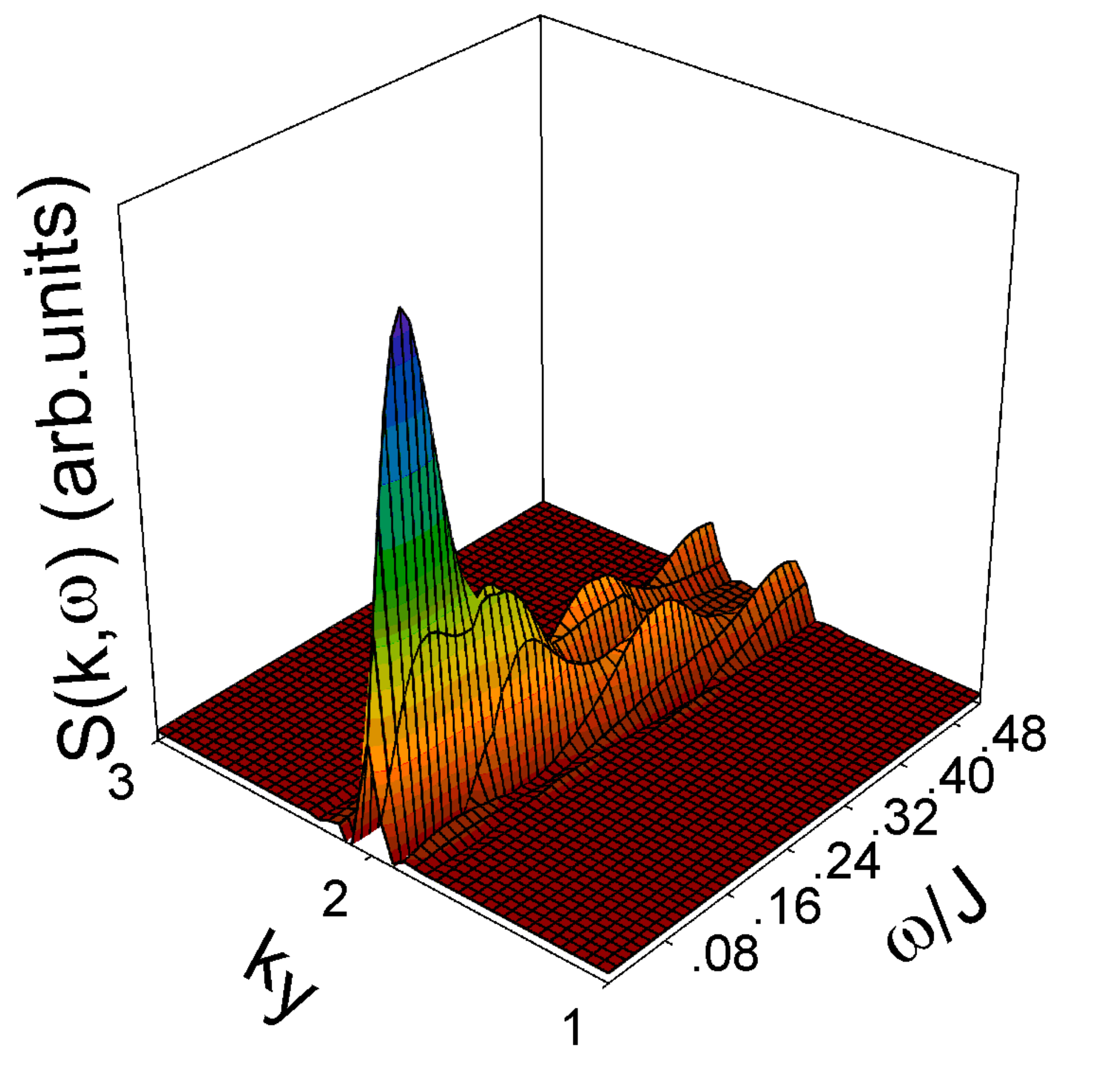}
\caption{(color online) The energy dependence of
dynamical spin structure factor $S({\bf k},\omega)$ along the direction $k_{y}$ ($k_{x}=\frac{2\pi}{3}$) in first BZ
at $x_{{\rm opt}}=0.29$.}
\end{figure}

We plot in Fig. 2 the dynamical spin structure factor $S({\bf k},\omega)$ of the electron-doped cobaltates NCO in the ($k_{x},k_{y}$) (here we use the unit of [$\pi/3,\pi/\sqrt{3}$]) plane at the optimal doping $x_{{\rm opt}}=0.29$ with temperature $T=0.002J$ for parameter $t/J=-3.0$ at energy (a) $\omega =0.08J$, and (b) $\omega =0.20J$. We find that a spin resonance peak locates at antiferromagnetic wave vector $Q (\frac{2\pi}{3},\frac{2\pi}{\sqrt{3}})$ at low energy, then evolves outward into six incommensurate magnetic scattering peaks at high energy.  Along one direction $k_{y}$ ($k_{x}=\frac{2\pi}{3}$), we calculate the energy dependence of $S({\bf k},\omega)$ and find in Fig. 3 that a commensurate spin resonance peak locating at antiferromagnetic wave vector $Q (\frac{2\pi}{3},\frac{2\pi}{\sqrt{3}})$ persists in a broad range of low energies up to about $0.16J\approx$3.2 mev, then splits into two incommensurate magnetic scattering peaks with increasing energies (The energy dependence of $S({\bf k},\omega)$ along the other two directions display the same transition feature and don't show here). The present result is very similar to the spin resonance excitation of electron-doped cupates\cite{wilson,feng3,ilya}, and differs from the case of hole doped cuprate in which another incommensurate-commensurate spin resonance excitations transition happens at low energies\cite{dai,feng1}. Therefore the commensurate magnetic resonance could be intimately related to superconductivity and is a common and essential feature in doped Mott insulators, other details such as the incommensurability and hour-glass dispersion found in different cuprates superconductors may not be fundamental to superconductivity.

\section{Summary and Discussions}

In summary, we investigate the unconventional superconductivity in electron-doped cobaltates NCO based on the mechanism of kinetic energy driven superconductivity. We find that the SC state is governed by both charge carrier pairing and quasiparticle coherent, and the resulting SC transition temperature is suppressed to a lower temperature due to the strong magnetic frustration, and shows a common doping-dependent dome-shaped feature, in agreement with experimental measurements. Then we theoretically analyze the feedback effect of superconductivity with $d_{1}+ id_{2}$-wave pairing symmetry on the magnetic excitations, and find a commensurate resonance peak locating at antiferromagnetic wave vector $Q (\frac{2\pi}{3},\frac{2\pi}{\sqrt{3}})$ for a broad range of low energies, then evolving outward into six incommensurate magnetic scattering peaks with increasing energy, strongly suggesting that magnetic resonance could be one of the fundamental features in doped Mott insulators. We propose that the inelastic neutron scattering experiment can verify the magnetic resonance excitations in electron-doped cobaltates NCO after the overcome of the difficulty in growing its high-quality SC sample.

\section{Acknowledgments}

This work was supported by NSFC under Grant No. 11104011, Research Funds of BJTU under Grants No. 2011JBM126 and No. 2011RC027, and National Basic Research Program of China under Grant No. 2011CBA00102. We thank helpful discussions with Prof. Shiping Feng.

\end{document}